# Field emission from self-catalyzed GaAs nanowires

Filippo Giubileo [1,\*], Antonio Di Bartolomeo [2,1], Laura Iemmo [2], Giuseppe Luongo [2,1], Maurizio Passacantando [3], Eero Koivusalo [4], Teemu V. Hakkarainen [4] and Mircea Guina [4]

[1] CNR-SPIN Salerno, via Giovanni Paolo II n.132, I-84084, Fisciano, Italy; filippo.giubileo@spin.cnr.it
[2] Physics Department 'E. R. Caianiello', University of Salerno, via Giovanni Paolo II, I-84084, Fisciano, Italy; adibartolomeo@unisa.it; liemmo@unisa.it; giluongo@unisa.it
[3] Department of Physical and Chemical Science, University of L'Aquila, via Vetoio, I-67100, Coppito, L'Aquila, Italy; maurizio.passacantando@aquila.infn.it
[4] Optoelectronics Research Centre, Tampere University of Technology, Korkeakoulunkatu 3, FI-33720 Tampere, Finland; eero.koivusalo@tut.fi; teemu.hakkarainen@tut.fi; Mircea.Guina@tut.fi
\* Correspondence: filippo.giubileo@spin.cnr.it; Tel.: +39-089-96-9329

**Abstract:** We report observation of field emission from self-catalyzed GaAs nanowires grown on Si (111). The measurements are realized inside a scanning electron microscope chamber with nano-controlled tungsten tip functioning as anode. Experimental data are analyzed in the framework of Fowler-Nordheim theory. We demonstrate stable current up to $10^{-7}$ A emitted from the tip of single nanowire, with field enhancement factor β up to 112 at anode-cathode distance d=350 nm. A linear dependence of β on the anode-cathode distance is experimentally found. We also show that the presence of a Ga catalyst droplet suppresses the emission of current from the nanowire tip. This allows detection of field emission from the nanowire sidewalls, which occurs with reduced field enhancement factor and stability. This study further extends the GaAs technology to vacuum electronics applications.

**Keywords:** Field emission; semiconductor nanowires; gallium arsenide; Fowler-Nordheim theory; field enhancement factor.

## 1. Introduction

Field emission (FE), that is the quantum mechanical tunneling of electrons from the material surface through the vacuum energy barrier when a sufficiently high electric field is applied, can be exploited for several applications in vacuum electronics, such as flat panel displays [1,2], electron [3] and x-ray sources [4], and microwave devices [5]. Nanostructures represent the best candidates to operate as field emitter sources due to the high aspect ratio that enables high local field enhancement. Several 1D and 2D carbon based nanostructures have been characterized as field emitters: Aligned carbon nanotube (CNT) films [6-8], single CNT [9,10], CNT networks [11-13], graphene [14-16], graphene oxide nanosheets [17]. One-dimensional semiconductor nanostructures, such as nanowires (NWs), nanorods, nanoparticles, etc., have also attracted considerable attention due to wide applicability for functional devices in the field of optoelectronics [18,19], photovoltaics [20,21] as well as vacuum electronics [22]. Several studies on NWs (GaN [23,24], ZnO [25], $W_5O_{14}$ [26]) and nanoparticles ($In_2O_3$ [27], GeSn [28]) have been reported. GaAs, which is one of the most popular III–V compound semiconductors with high electron mobility and direct band gap, in the form of nanowires (NWs) can have interesting properties for FE applications. A particularly interesting III-V nanomaterial system are the self-catalyzed GaAs NWs grown by vapor-liquid-solid method [29,30] allowing direct integration of high quality GaAs structures on Si without the use of Au or other foreign catalyst metals, which would introduce deep level traps in Si [31]. Despite that, very limited research on FE from nanostructured GaAs is available to date.  Porous GaAs having closely spaced nanometric crystallites [32], obtained by anodic etching of *n*-type (110) GaAs, was demonstrated as a weak (10 nA for 4 kV applied bias) and unstable emitter with large current fluctuations and surface modifications within few hours. 1D pillars fabricated by electrochemical etching of (111) GaAs substrate [33,34] resulted too thick (diameter ~2μm) for FE applications. Better performance were reported for GaAs nanowires fabricated by electrochemical etching of anodic etched *n*-type GaAs (111) wafer [35]. Nanowires were actually aggregated as bundles with an average top diameter in the range 30 to 80 μm. Experimental data showed that such bundles work as field emitters with low turn-on field ($E_{ON}$ ~ 3V/μm). Similar turn-on field (~2V/μm) was measured in parallel plate configuration (sample area 40 mm$^2$) for aligned GaAs NWs fabricated by direct etching by H plasma of the GaAs wafer covered with Au film [36]. However, a systematic study of field emission from a single GaAs NW or from an array GaAs NWs is still missing.

In this Letter we characterize the field emission properties of self-catalyzed GaAs nanowires, fabricated with a lithography-free method by self-catalyzed growth on Si/SiO$_x$ patterns. The effect of *n*-doping as well as the influence of Ga droplets on the top of nanowires are studied. We report stable emitted current from GaAs nanowires, with field enhancement factor up to $\beta$ =112 at anode-cathode separation of 350 nm for highly n-doped samples. Taking advantage from the suppression of field emission by Ga droplets on the NW tips, we also report emission from the NW sidewalls, although with lower field enhancement factor and limited current stability.

## 2. Experimental

Self-catalyzed GaAs NWs were grown on Si(111) substrate by molecular beam epitaxy using a droplet epitaxy method [37,38] to form nucleation sites (oxide-free areas) on the substrate, with control of size and density of the sites. Ga catalyst droplets were formed in such sites, and GaAs NWs growth was obtained by simultaneous deposition of Ga and As. The Ga catalyst droplets were either preserved or removed by crystallizing them into GaAs in As flux after the NW growth. In this experiment, we measured three samples: Sample-1 consists of an array of NWs with ~143 nm diameter and 4×10$^7$ cm$^{-2}$ density and with Ga droplet on the tips; Sample-2 has NWs with ~130nm diameter, 6×10$^7$ cm$^{-2}$ density and Ga-droplet free tip; in Sample-3, the NWs have diameter and density of ~192 nm and 6×10$^7$ cm$^{-2}$, respectively, are covered by Ga droplets and are n-doped by Te with nominal doping 2×10$^{19}$ cm$^{-3}$.

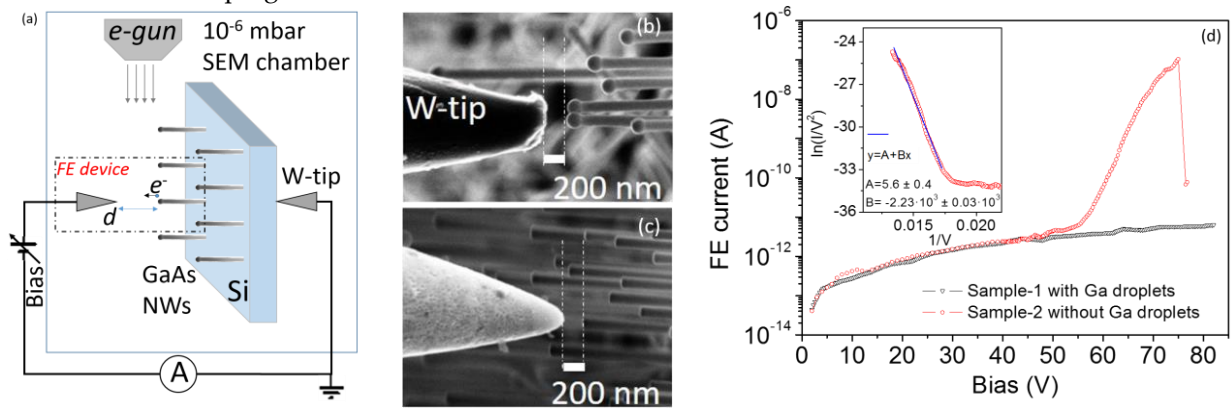

**Figure 1.** (**a**) Field emission setup realized inside a SEM chamber by using two nano-manipulated tungsten tips. Image of FE device with the W-tip at d = 200 nm from the NWs of (**b**) Sample-1 (NW tips covered by Ga droplets) and (**c**) Sample-2 (free NW Tips). The SEM sample stage was rotated to allow precise estimation of tip-sample distance; (**d**) FE current-voltage characteristics measured in the voltage range 0-80V for both samples. Inset: Fowler-Nordheim plot $\ln(I/V^2)$ vs V showing linear behavior $y = A + Bx$ with $B = (-2.23 \pm 0.03) \times 10^3$ and $A = (5.6 \pm 0.4)$, to confirm the FE nature of the measured current for Sample-2.

Field emission measurements were performed at 10$^{-6}$ mbar base pressure inside the vacuum chamber of a Zeiss LEO 1530 field emission scanning electron microscope (SEM), equipped with

Kleindiek piezo-controlled nanomanipulators. The two probes (tungsten tips) were used as electrodes, the cathode contacting the sample, the anode being positioned in front of the NWs at a controlled separation $d$ to collect the emitted electrons. SEM stage was tilted respect to the electron beam to acquire cross-sectional image, in order to favor the estimation of the tip-sample distance. A semiconductor parameter analyzer (Keithley 4200 SCS) was used as source-meter unit to apply bias (in the range 0 – 100 V) and to measure the current from the FE device with resolution better than 1pA. A schematic of the experimental setup is reported in Figure 1a.

## 3. Results and discussion

FE measurements were realized by gently approaching the anode-tip close to a NW apex. In order to check the effect of Ga droplets on the GaAs NW tips, we compared the current-voltage ($I - V$) characteristics measured, at the same separation $d = 200$ nm, on Sample-1 with Ga droplets (figure 1b) and on Sample-2 without Ga droplets (figure 1c), respectively. The emission currents are reported in figure 1d. Despite a high applied voltage in the range 0-80 V, we found that in Sample-1 the presence of Ga droplets inhibits the emission of electrons from the NWs. On the other hand, for Sample-2, a rapidly increasing current is measured for bias above 45 V. The turn-on field $E_{ON} = V_{ON}/d$ is ~ 0.22 V/nm being defined here as the field necessary to achieve a current of $10^{-11}$ A. Considering that the anode is tip-shaped (differently from the most common parallel plate geometry), a more accurate estimation of the turn-on field can be obtained by including a tip correction factor [7] $k \approx 1.5$, which yields a lower turn-on field $\tilde{E}_{ON} = E_{ON}/k \approx 0.15$ V/nm. The relative high turn-on field can be explained by the very small cathode-anode separation. Indeed, as demonstrated for CNTs, the turn-on field is strongly dependent on the electrode distance [39] with a reduction of the field value about ten times while increasing the separation from 1 to 60 µm. According to the Fowler-Nordheim theory [40], the FE current $I$ can be expressed as a function of the applied bias $V$ as follows:

$$I = S \cdot a \frac{\beta^2 V^2}{\varphi d^2} exp\left(-b \; d \frac{\varphi^{3/2}}{\beta V}\right) \qquad (1)$$

where $a = 1.54 \times 10^{-6} \; A \; V^{-2} eV$ and $b = 6.83 \times 10^9 \; eV^{-3/2} m^{-1} V$ are constants, $S$ is the emitting area, $\beta$ is the field enhancement factor taking into account the field amplification at an apex, and $\varphi$ is the workfunction of the GaAs NW. From this expression it is immediately verified that by plotting $ln(I/V^2)$ vs $1/V$, the so-called Fowler-Nordheim (FN) plot, a linear behavior is expected:

$$ln\left(\frac{I}{V^2}\right) = \; m \cdot \frac{1}{V} + y_0 \qquad (2)$$

where the slope is $m = -bd\varphi^{3/2}/\beta$ and the intercept is $y_0 = ln(S \cdot a \beta^2/(\varphi d^2))$. Actually, this is a standard procedure used to confirm a FE phenomenon also for nanostructured emitters, although the FN theory was developed for a flat conductor. However, it is widely accepted that it can be properly applied to nanostructures with good approximation [41], the field enhancement factor β taking into account the amplification occurring around an apex. The current, that appears very stable without particular fluctuations, increases by more than four orders of magnitude (from $10^{-11}$A to $10^{-7}$A) in the bias range from 45 V to 80 V (figure 1d). At 80 V a dramatic modification of the FE device happens with the evaporation of the NW from the substrate and the interruption of the emitted current. We clarify here, that despite the high number of NWs on the surface, considering the sharp tungsten tip (curvature radius ~100 nm) and the average density of NWs ($4 \times 10^7$ cm$^{-2}$), corresponding to an average spacing between the NWs of ~1 µm, we typically obtain FE devices in which only one NW contributes to the emitted current. As a matter of fact, we systematically recorded about the same maximum current, which is likely the highest current that a single undoped NW can sustain (~$10^{-7}$ A). In the inset of figure 1d we show the FN-plot: Data are very well fitted by a straight line confirming the FE nature of the observed current. From the slope and the intercept of the fit line, assuming $\varphi = 4.77$ eV for the workfunction of GaAs [42], we extract a field enhancement factor $\beta \approx 7$. We notice that lower turn-on field ($E_{ON}$ = 2.0 V/µm) and higher field enhancement factor ($\beta$ = 3500) have been reported [36] for high-density aligned GaAs nanowires (40 mm² sample area) measured in the parallel plate configuration with 5 cm diameter anode and applied voltage up to 8 kV at a separation up to 4 cm. However, we have to take into account that these parameters ($\beta$ and

$E_{ON}$) depend on the inter-electrode distance as well as on the aspect ratio and the spatial distribution of the emitters. Moreover, they strongly depend on the setup used for FE measurements. The parallel plate setup studies FE current averaged over an enormous number of emitters, while tip anode setup probes the emitters individually. A consistent comparison of reported values is quite a complex task unless similar experimental conditions are adopted. Indeed, it has been demonstrated that when realizing small FE devices with micro- or nano-sized metallic tip as collector electrode, the turn-on field (field enhancement factor) is strongly increased (reduced) due to the quantum screening effect that is detrimental for the FE performance [43].

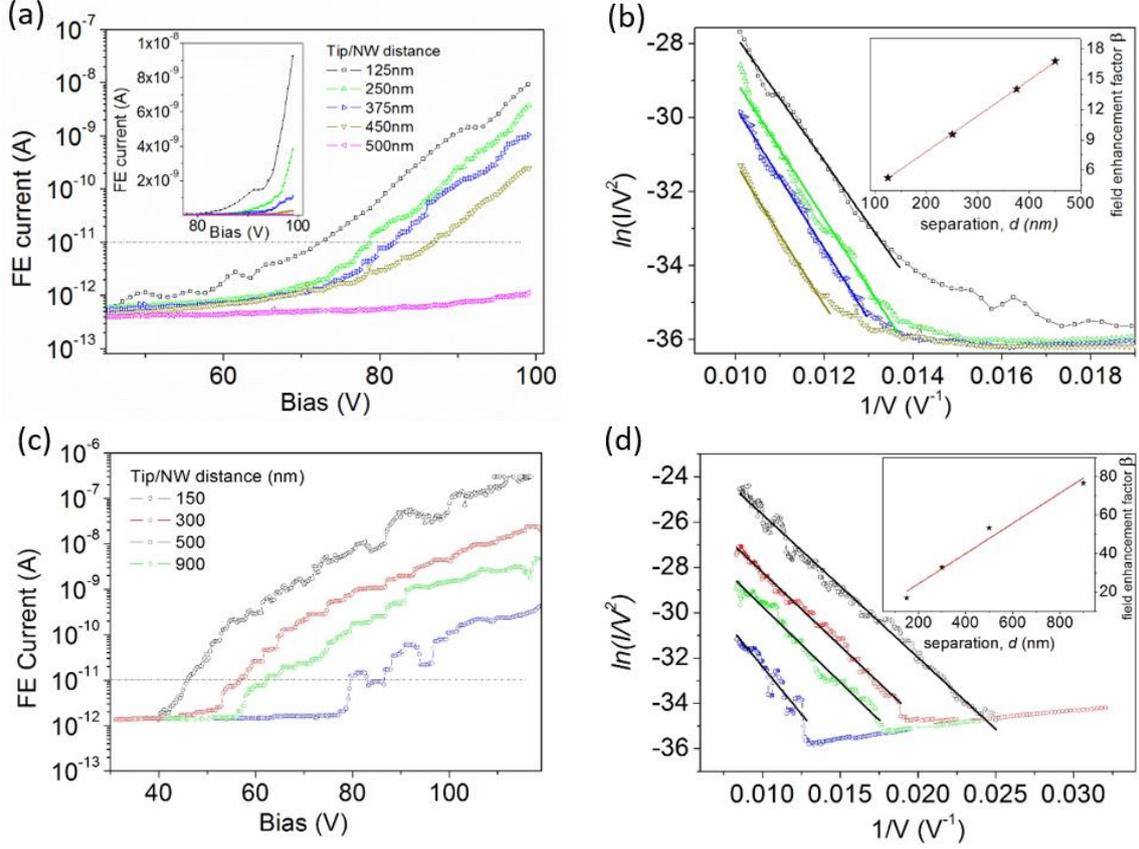

**Figure 2.** FE characterization of Sample-2 without Ga droplets. (**a**) Semi-log plot of the $I-V$ characteristics measured for Sample-2 for different values of the separation $d$. Dotted line identifies the current level at which we define the turn-on field $E_{ON}$. In the inset the characteristics are reported in linear scale; (**b**) FN-plots and linear fittings (solid lines). From the slope of fitting lines we extracted the field enhancement factor $\beta$, plotted as a function of $d$ in the inset; (**c**) Semi-log plot of the $I-V$ characteristics measured for Sample-2 for different values of the separation $d$ in a different location of the sample; (**d**) FN-plots and linear fittings (solid lines). Inset: $\beta$ vs $d$.

As a confirmation of the dependence of $\beta$ as an increasing function of distance, we show in figure 2 the evolution of the $I-V$ characteristics by varying the separation $d$ between the tungsten tip (anode) and the apex of a GaAs NW on Sample-2. In figure 2a we show the recorded $I-V$ curves for tip-sample separation in the range 125 nm to 500 nm. We clearly observed a rapid increase of the current above 70 V. The corresponding FN plots reported in figure 2b confirm the FE nature of the measured current. From the slope of the plots we extracted the field enhancement factor for each value of $d$ and the inset show the experimental evidence of a linear dependence of $\beta$ vs $d$. According to this behavior, we can consider the $\beta$ value rather high taking into account the small separation distance. A similar behavior has been actually measured in several different locations on Sample-2. As an example we report in figure 2c another set of measurements performed in a different location by varying the separation $d$ in the range 150 nm to 900 nm. We observe that in both cases the emitted current raises from the noise-floor level ($10^{-12}$A) for at least four orders of magnitude in a voltage range about 40-50V wide. Again we extracted a clear linear dependence of $\beta$ vs $d$ confirming that $\beta$ increases with $d$. The difference on the absolute value of $\beta$ is easily understood by considering that many factors may influence it, such as the length of NW as well as small local variations of the workfunction.

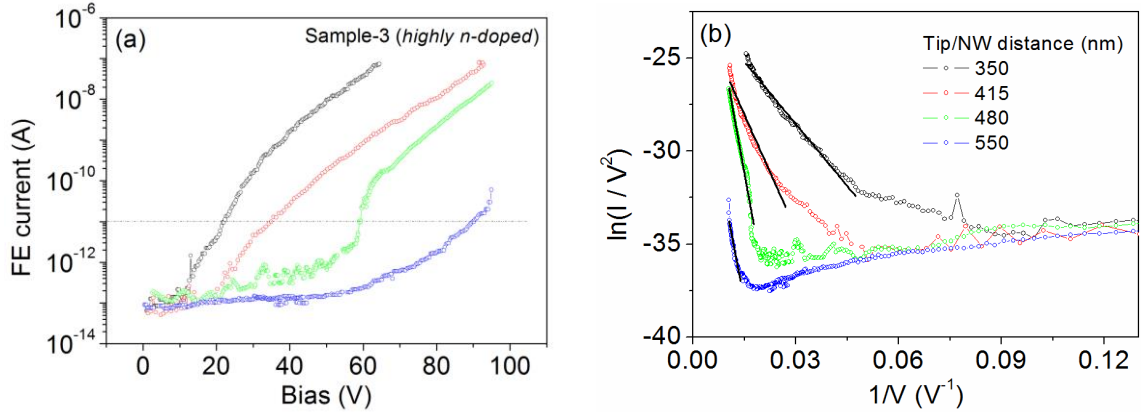

**Figure 3.** FE characterization of highly n-doped GaAs NWs. (**a**) $I-V$ characteristics measured for Sample-3 for different values of the separation $d$. (**b**) FN-plots and linear fittings (solid lines).

From a theoretical point of view, the field enhancement factor could be estimated by considering the simplified model by Edgcombe and Valdrè [44] for a cylindrical emitter of height $h$ and a semi-spherical apex with radius $r$. According to this model, the field enhancement factor is expected to be $\beta = 1.2 \times (2.15 + h/r)^{0.9} \approx 45$ for a single GaAs NW emitter in Sample-2 (height $h \cong 3500$ nm and $r \cong 65$ nm). According to this relation, it results that variations of the NW height correspond to variations in $\beta$. In case the separation distance $d$ between the NW apex and the anode (tungsten tip) is small, i.e. $d \leq 0.3h$, a further increase of $\beta$ is expected [45,46] according to the formula

$$\beta = 1.2 \times \left(2.15 + \frac{h}{r}\right)^{0.9} \left[1 + 0.013 \left(\frac{d}{d+h}\right)^{-1} - 0.033 \left(\frac{d}{d+h}\right)\right], \qquad (3)$$

from which we found out $\beta \approx 60$ for $d = 150$ nm. The expected values are in good agreement with the values we extracted from our experimental data. The lower values obtained in some cases can be understood by taking into account the presence of neighbor NWs can produce a significant screening effect depending on the relative spacing $s$. Indeed, it has been demonstrated that for vertically aligned tubes [47,48] the field enhancement factor depends on the spacing $s$, and it can be expressed as $\tilde{\beta} = \beta(1 - e^{-2.31 \cdot s/h})$. Consequently, the spacing s has a crucial impact on the measured $\beta$ value. In our sample, the average spacing is about 1μm or below, and corresponds to a range of spacing in which the field enhancement factor is rapidly changing with $s$. Consequently, when probing a single NW on the sample, the extracted $\beta$ can be affected by the actual distribution of the NWs in the neighborhood. This information gives clear indication that in order to realize highly uniform large area emitting surface is necessary to fabricate ordered arrays of NWs.

We also characterized the highly n-doped GaAs NWs (Sample-3) in which Te atoms have been introduced to obtain a nominal doping of $2 \times 10^{19}$. Although the Ga droplets were not intentionally removed in the process of Sample-3, statistically significant number of NWs were found to be droplet-free. Our special setup for FE measurements, being inside a SEM chamber, allowed to select those NWs without Ga droplets to characterize the FE from apex. Experimental data are reported in figure 3. It is immediately evident from the $I-V$ characteristics (figure 3a) that, although we are working in a similar range of tip-NW separation, the turn-on voltage is significantly lower (~20V) corresponding to a turn-on field of 0.057 V/nm. From the linear fit of FN-plots (figure 3b) we can extract the field enhancement factor. For the minimum distance (*d*=350 nm) we obtain the highest factor $\beta \approx 112$, with respect the values extracted for the undoped samples. This result is not surprising: Te (group VI element) is expected to produce n-type doping in GaAs NWs [49]. The doping atoms modify the electronic structures of nanowires by introducing donor states causing higher local electron states near the Fermi level. Consequently, more electrons which can tunnel (at given voltage) through the barrier to vacuum are provided and Fermi level is moved to near vacuum level (decrease of the work function). Moreover, chemical doping has been often used to improve FE properties in several nanostructures such as CNTs [50-52], TiO2 nanotubes [53], GaN NWs [54], etc.

Concerning Sample-1, we have shown that Ga droplets prevents the field emission from the NW apex. On the other hand, the workfunction of Ga is 4.2 eV, and therefore the suppression of field emission can be explained considering that air exposed Ga is oxidized and acts as an extra dielectric layer. However, this condition opens the opportunity to check the emission from the (110)-facetted sidewalls of the hexagonal NWs. The $I-V$ characteristics (figure 4a) have been measured by allowing the tungsten tip (anode) to translate parallel to the NW axis but shifted (about 1 μm) on a side of the NW (figure 4b-e). If the anode is above the Ga droplet no current is recorded (figure 4b). As soon as the tip apex goes beyond the droplet a small current appears (figure 4c). Further forward steps, which increase the axial overlap of the tip with the NW (figure 4d-f), result in higher current laterally emitted from the NW. The F-N plots reported in the inset are linear and they confirm the FE nature of the measured current. From a quantitative point of view, to extract the field enhancement factor from such experimental data is a quite complex task due to the setup configuration (that does not allow precise estimation of the tip-NW separation). However, in order to compare the FE properties of the lateral surface with the NW apex, we can extract the ratio $\beta/d$ from the slope of the FN-plot. We found that from our experimental data 0.01 nm$^{-1}$ < $\beta/d$ < 0.04 nm$^{-1}$ for the lateral emission while 0.04 nm$^{-1}$ < $\beta/d$ < 0.08 nm$^{-1}$ for the apex emission, that confirms a better FE performance for the latter case.

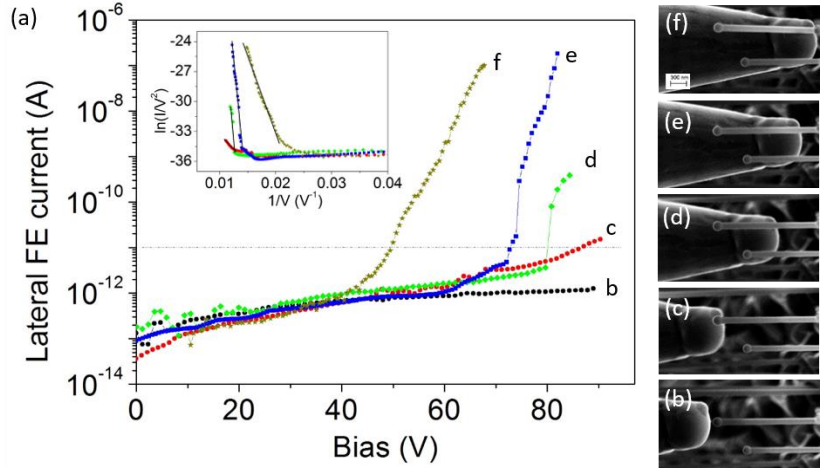

**Figure 4.** (**a**) $I-V$ characteristics measured in the lateral FE configuration. Curves refer to the SEM images (**b**), (**c**), (**d**), (**e**), (**f**). Inset: FN-plots and linear fits.

Finally, one important issue in FE characterization is the emission stability and lifetime. Previous works on GaAs based field emitters [32,35] report very unstable current emission vs time. We applied a constant voltage and recorded the emission current versus time $I(t)$, with a sampling time of 1 s. In figure 5 we show the current variation as a function of time for the lateral emission from NW with Ga droplets (figure 5a), for the undoped NW without Ga droplets (figure 5b) and for the highly n-doped NW (figure 5c). The samples showed very stable behavior, with constant emitted current, without evident degradation for a testing operational time of 1 h. Statistical analysis of the measured current values is reported for Sample-1 (figure 5d), Sample-2 (figure 5e) and for Sample-3 (figure 5f). A very good stability was obtained on a time period of about 1 h for Sample-2 and Sample-3 (current emitted from the NW apex), with less than 20% deviation from the average current $I_{Mean}$. On the contrary, larger fluctuations were recorded for the lateral emitted current. The reduced stability in the case of lateral emission can be related to the different location of the anode (parallel to the sidewall). The application of high electric field can bend the nanowire towards the other electrode due to the electrostatic attraction and this results in the observed instability.

The high current stability versus time can be considered a very good result compared to reported instabilities and it confirms that high quality aligned GaAs nanowires are suitable for long operational FE devices.

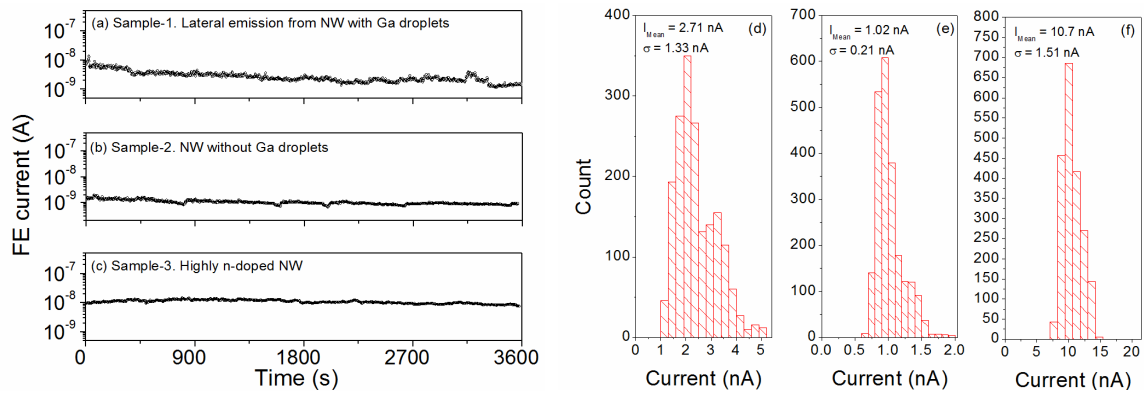

**Figure 5.** Comparison of the current stability (FE current vs time) for Sample-1 (**a**), measured at constant bias of 60 V, for Sample-2 (**b**) measured at constant bias of 90 V, and for Sample-3 (**c**), measured at constant bias of 70 V. Histograms to summarize the statistical analysis on the current values are reported for Sample-1 (**d**), for Sample-2 (e) and for Sample-3 (f).

## 4. Conclusions

In conclusion, we have extensively analyzed the field emission properties of self-catalyzed GaAs nanowires on Si(111) grown on Si. We compared FE performance between undoped and highly *n*-doped NWs, with the highest field enhancement factor of 112 recorded at the small separation distance *d*=350 nm for doped NWs. We explain the observation of different $\beta$ values on the same sample as the effect of the spacing between NWs. We demonstrate that oxidized Ga droplets at the NW apex are detrimental to FE phenomenon from the apex and we characterize the FE from the lateral surface of the NWs, estimating reduced performance parameters with respect apex emission. Finally, we prove high current stability vs time, with average fluctuations below 20%, which is a prerequisite for device exploitation.

**Acknowledgments:** Funding from the Academy of Finland Project NESP (decision number 294630) is acknowledged by E.K., T.V.H. and M.G. Funding from the Regione Campania, Legge 5 bando 2008 prot. 2014 0293185 29/4/2014 is acknowledged by A.D.B.

**Author Contributions:** Author Contributions: F.G., A.D.B. and M.P. conceived and designed the experiments. E.K., T.V.H and M.G. fabricated and characterized GaAs nanowires. F.G., G.L., and M.P. performed field emission experiment. F.G., L.I. and A.D.B. analyzed field emission data. All authors discussed the results, contributed to the manuscript text, commented on the manuscript, and approved its final version.

**Conflicts of Interest:** The authors declare no conflict of interest.